\documentclass[aip,graphicx,reprint]{revtex4-1}

\usepackage{amssymb}
\usepackage{graphicx}

\begin{document}


\title{Thresholdless nonlinearity-induced edge solitons in trimer arrays} 



\author{Magnus Johansson}
\email[]{magjo23@liu.se}
\affiliation{Department of Physics, Chemistry and Biology (IFM), Linköping University, SE-581 83 Linköping, Sweden}


\date{\today}

\begin{abstract}
  We consider a one-dimensional discrete nonlinear Schrödinger (DNLS) model with
  Kerr-type on-site nonlinearity, where the nearest-neighbor coupling
  constants take two different values ordered in a three-periodic sequence.
  The existence of localized edge states in the linear limit
  (Su-Schrieffer-Heeger trimer, SSH3)
  is known to depend on the precise location of the edge. Here, we show
  that for a termination that does not support linear edge states,
  an arbitrarily weak on-site nonlinearity will induce an edge mode with
  asymptotic exponential localization, bifurcating from a linear band edge.
  Close to the gap edge, the shape of the mode can be analytically described
  in a continuum approximation as one half of a standard gap soliton. The
  linear stability properties of nonlinear edge modes are also discussed. 
\end{abstract}


\maketitle 

\section{Introduction}
\label{sec:Intro}
Recent years have shown a large burst of activity related to topological
properties in various physical setups, and in particular within the field
of photonics where nonlinear effects often play an important role. For reviews,
see, e.g., Refs.\ \onlinecite{SLCK20,AC22,Kruk23,SR24}.
As a standard one-dimensional
model displaying topological features in the linear regime,
such as bulk-edge correspondence between existence of edge modes and a
nontrivial winding number (Zak phase) of Bloch bands, the
Su-Schrieffer-Heeger (SSH) model\cite{SSH79} is frequently considered.
In its simplest
version it describes a tight-binding lattice without on-site potential
and with binary alternating nearest-neighbor coupling constants, and many works,
theoretical as well as experimental, considered the effects of including
nonlinear effects of various forms on the topological properties of the SSH
model. See, e.g., Ref.\ \onlinecite{Bugarski} and references therein for
recent results on the SSH model with cubic (Kerr) or saturable nonlinearities.

An extension of the SSH model that has received increasing attention during
the last few years is the SSH trimer (SSH3) model, with a three-site unit cell
connected by three alternating coupling constants. In the linear case, several
recent works\cite{Alvarez,Diakonos22,Yang24}
have described properties of edge states and proposed various
formulations of a bulk-edge correspondence which, for the most general trimer
configurations, is less evident than for the SSH dimer due to absence of
mirror-symmetry. Very recently, an acoustic analog of the SSH3 lattices
was also proposed, and conditions for existence of edge modes obtained
and confirmed experimentally\cite{Diakonos24}. As is known from the above
works, the existence of edge states in the SSH3 model
depends crucially on the relation between
the coupling constants as well as on the location of the edge; a phase
diagram for the general case with three different coupling constants was
illustrated in Fig.\ 6 of Ref.\ \onlinecite{Diakonos22}.

A particular configuration of a trimer array was also studied in the
nonlinear regime by Kartashov et al.\cite{Kartashov22} and, using an
experimental realization with fs-laser written waveguide arrays, edge solitons
were confirmed to bifurcate from linear edge modes in both of the topological
gaps of the linear trimer model. Importantly, the implementations studied in
Ref.\ \onlinecite{Kartashov22} all started with two equal coupling constants
closest
to the edges, where two linear edge modes exist at each edge only if these two
coupling constants are weaker (i.e., with a weakly coupled trimer located
at the edge). Qualitatively, the continuations of the linear gap edge modes
into nonlinear edge modes were seen to exhibit regimes of instabilities and
bifurcations similar to those of the dimer SSH model\cite{MS21,Bugarski}.

However, if one considers the same bulk trimer configuration as in
Ref.\ \onlinecite{Kartashov22} but instead terminates the chain with the single
strong bond (see Fig.\ \ref{fig:trimer} below),
a more intricate situation arises. From the phase diagram in
Ref.\ \onlinecite{Diakonos22}
this is seen to correspond to a borderline case where
no localized linear edge state exists, but will be created by an arbitrarily
weak symmetry breaking if the third bond becomes slightly stronger than the
second. It is thus interesting to find out what happens for this
configuration if the symmetry between the two weak bonds is preserved,
but instead an on-site Kerr nonlinearity added. As we will show here, indeed an
arbitrarily weak nonlinearity is sufficient to induce a localized
nonlinear edge mode, with frequency bifurcating from the edge of a linear
band. Close to the linear band,  the edge mode can be described
in a continuum approximation using an approach analogous to that used
already 30 years ago by Usatenko, Kovalev and Vyalov\cite{UKV95}, who then
considered bulk gap solitons in three-atomic elastic chains with anharmonic
nearest-neighbor interactions. The small-amplitude edge mode in the
nonlinear SSH3 system
will take the shape of a standard gap soliton, cut by the edge at its center.
This is fundamentally different from the standard SSH model, where nonlinear
edge modes bifurcate from the localized linear edge mode at the gap center,
and in the
continuous limit only consists of a portion of the tail of the corresponding
bulk gap soliton\cite{SSLK19,MS21}.

\section{The nonlinear trimer model and its linear dispersion relation}
\label{sec:model}
 \begin{figure}[h]
 \includegraphics[width=0.49\textwidth,angle=0]{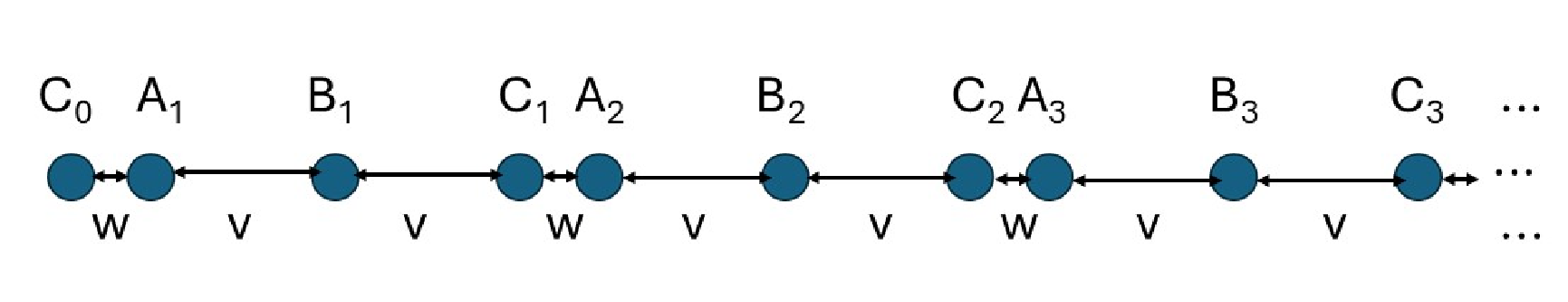}
\caption{The configuration of the semi-infinite trimer array.}
  \label{fig:trimer}
 \end{figure}
 The semi-infinite trimer array illustrated in Fig.\ \ref{fig:trimer} has a
 three-site unit
 cell. Labelling site indices with $j=1,2,...$ and cell indices with
 $n=0,1,2,..$, the
 following notation for the site amplitudes $\psi_j$ is used
 (the reason for this choice will become clear later),
 \begin{eqnarray}
   \psi_{3n}=B_n,
   \nonumber \\ 
   \psi_{3n+1}=C_n,
   \nonumber \\
   \psi_{3n+2}=A_{n+1},
   \label{eq:ABC}
 \end{eqnarray}
 with the left edge defined by the boundary condition $B_0=0$.

 The equations of motion for the SSH3 model with on-site
 (focusing) Kerr nonlinearity and positive coupling constants $v,w$
 can then
 be written in the standard form
 \begin{eqnarray}
   i \dot{A}_n+wC_{n-1}+vB_n+|A_n|^2 A_n = 0 ; n \geq 1,
   \nonumber \\
   i \dot{B}_n+v(A_{n}+C_n)+|B_n|^2 B_n = 0 ; n \geq 1 ,
   \nonumber \\
   i \dot{C}_n+vB_{n}+wA_{n+1}+|C_n|^2 C_n = 0 ; n \geq 0 .
   \label{eq:dyn}
    \end{eqnarray}
 
 The linear spectral bands are obtained in the standard way by
 neglecting the nonlinear terms and replacing the
 edge with periodic boundary conditions, $B_0=B_N, C_N=C_0$. From the ansatz
 $$(A_n,B_n,C_n)=(A,B,C)e^{i(kn+\lambda t)} ,$$
 the linear dispersion relation $\lambda(k)$ is determined by the solutions
 to the cubic equation
 \begin{equation}
   \lambda^3-(w^2+2v^2)\lambda -2v^2w \cos(k) = 0, 
   \label{eq:disp}
   \end{equation}
 illustrated in Fig.\ \ref{fig:disp}. 
  \begin{figure}[h]
 \includegraphics[width=0.4\textwidth]{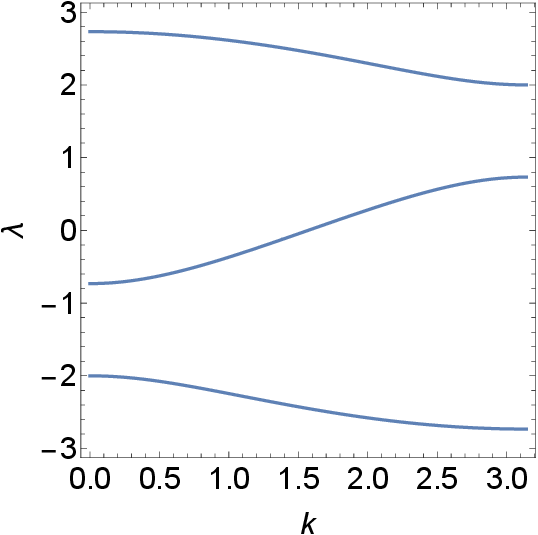}
 \caption{Dispersion relation (\ref{eq:disp}) with $w=2$, $v=1$.
   \label{fig:disp}} 
 \end{figure}
Note that  $\lambda(k)=-\lambda(\pi-k)$,
with the nonzero eigenvalues at the band center given by
$\lambda_{\pm}(k=\pi/2)=\pm\sqrt{w^2+2v^2}$. The eigenvalues at the band edges
also have simple expressions,
$\lambda_-(k=0)=-w$, $\lambda_0(k=0)=(w-\sqrt{w^2+8v^2})/2$,
$\lambda_+(k=0)=(w+\sqrt{w^2+8v^2})/2$. (We here assumed $w>v$, if $w<v$
the eigenvalues $\lambda_-$ and $\lambda_0$ will switch order.)

The linear eigenmode corresponding to the upper edge of the lower band
($k=0$), with
$\lambda=-w$, has the simple structure $A=-C$, $B=0$. Thus, this eigenmode
is also an exact eigenmode for the semi-infinite chain, and, as we will
see below, will yield a localized edge mode with frequency entering the gap
in presence of focusing nonlinearity. (Due to the symmetry of the spectrum
the same scenario, apart from sign changes, will occur at the lower edge
of the upper band if the nonlinearity is defocusing.)

\section{The nonlinearity-induced edge mode}
\label{sec:edge}

\subsection{Limit of uncoupled dimers}
\label{sec:dimers}
The standard way to construct families of gap modes in various types of
nonlinear lattice models consists essentially of three steps: (i) find
an ``anticontinuous'' limit where the system is decoupled and can be solved
exactly, and determine an appropriate configuration of its time-periodic
solutions with frequency inside the gap of the linear spectrum; (ii) continue
the anticontinuous solution for increasing coupling between the
units; and (iii) continue the solution for fixed nonzero coupling versus
frequency in both directions, to determine whether it reaches the gap edges
or bifurcates with some other gap mode inside the gap. In this way,
complete families of fundamental bulk gap modes connecting the anticontinuous
solutions smoothly to the continuous gap solitons at one band edge, and to
nonlocalized ``outgap'' solitons at the other edge, have been identified e.g.
for the DNLS model with binary modulated on-site potentials\cite{GJ04}, and
for the nonlinear SSH model\cite{VJ09}.

Here, the structure of the linear mode at the upper edge of the lower band
suggests an appropriate anticontinuous limit $v=0$, which turns the system
into a chain of uncoupled dimers with internal coupling $w$. This is identical
to the uncoupled limit for bulk gap solitons in the standard SSH
model\cite{VJ09}, and the relevant solution for an edge mode in this limit can
be chosen as $C_0=-A_1=\sqrt{w+\lambda}e^{i\lambda t}$ with stationary frequency
$-w < \lambda < 0$, and all other sites at zero amplitude. An example
of a family of nonlinear edge modes resulting from continuing this solution
for increasing $v$ is shown in Fig.\ \ref{fig:vcont}. Analogously to the
scenario for the standard nonlinear SSH model\cite{VJ09,Bugarski}, the
mode remains localized close to the edge until the coupling $v$ reaches
a value where the frequency enters the band above the gap
(at $v/w=\sqrt{3/8}=0.612...$ in Fig.\ \ref{fig:vcont}), where it
develops a non-decaying oscillatory tail.
 \begin{figure}[h]
   \includegraphics[height=0.49\textwidth,angle=270]{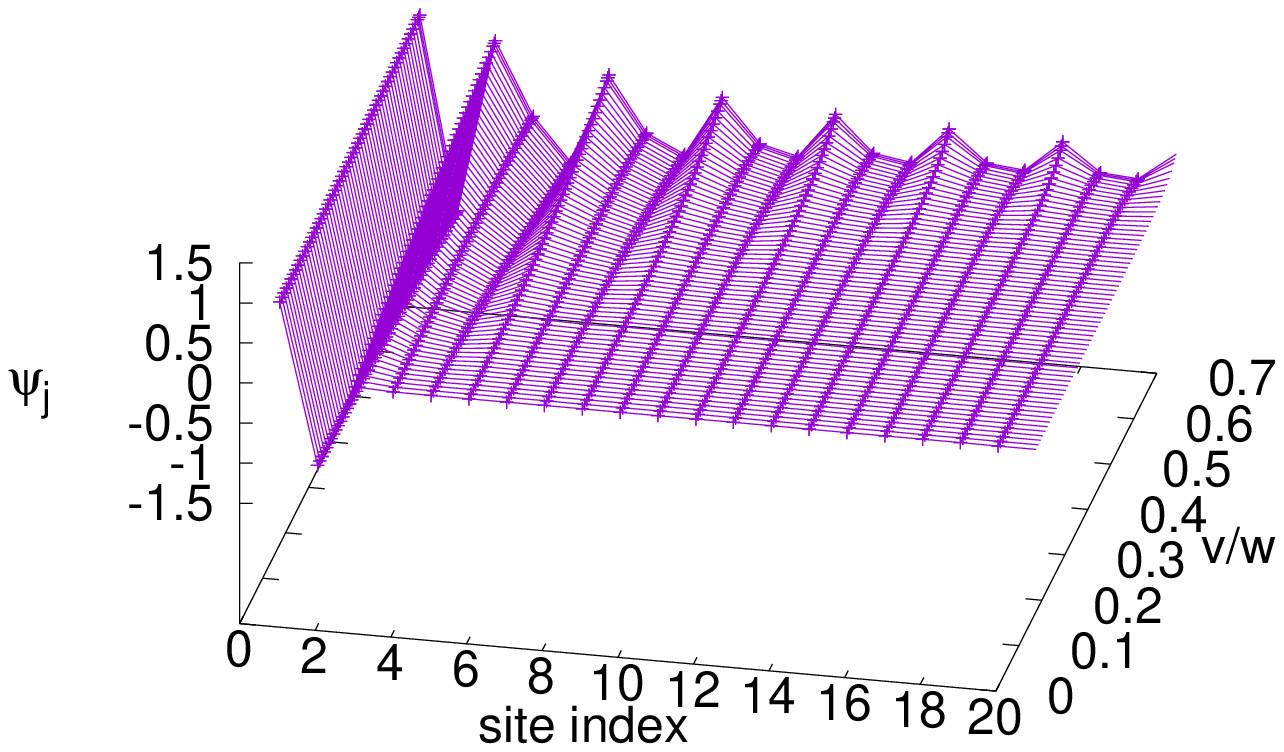}
   \caption{Continuation of the nonlinear edge mode versus $v$ from the
     limit $v=0$ of uncoupled dimers, with fixed $w=2$ and $\lambda=-1$. Only
     the part closest to the edge is shown.
   \label{fig:vcont}} 
 \end{figure}

\begin{figure}
   \includegraphics[height=0.49\textwidth,angle=270]{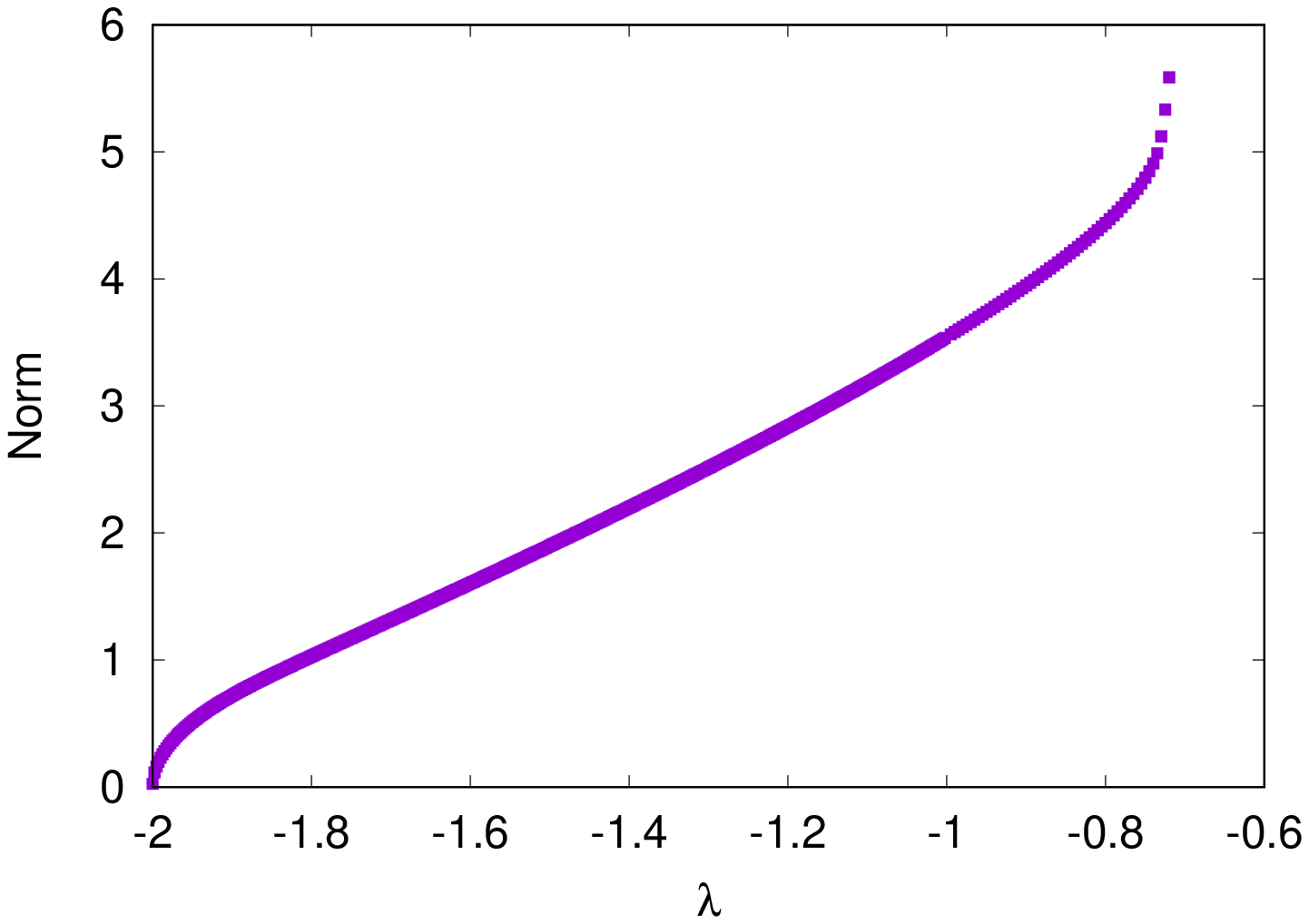}
 \includegraphics[height=0.49\textwidth,angle=270]{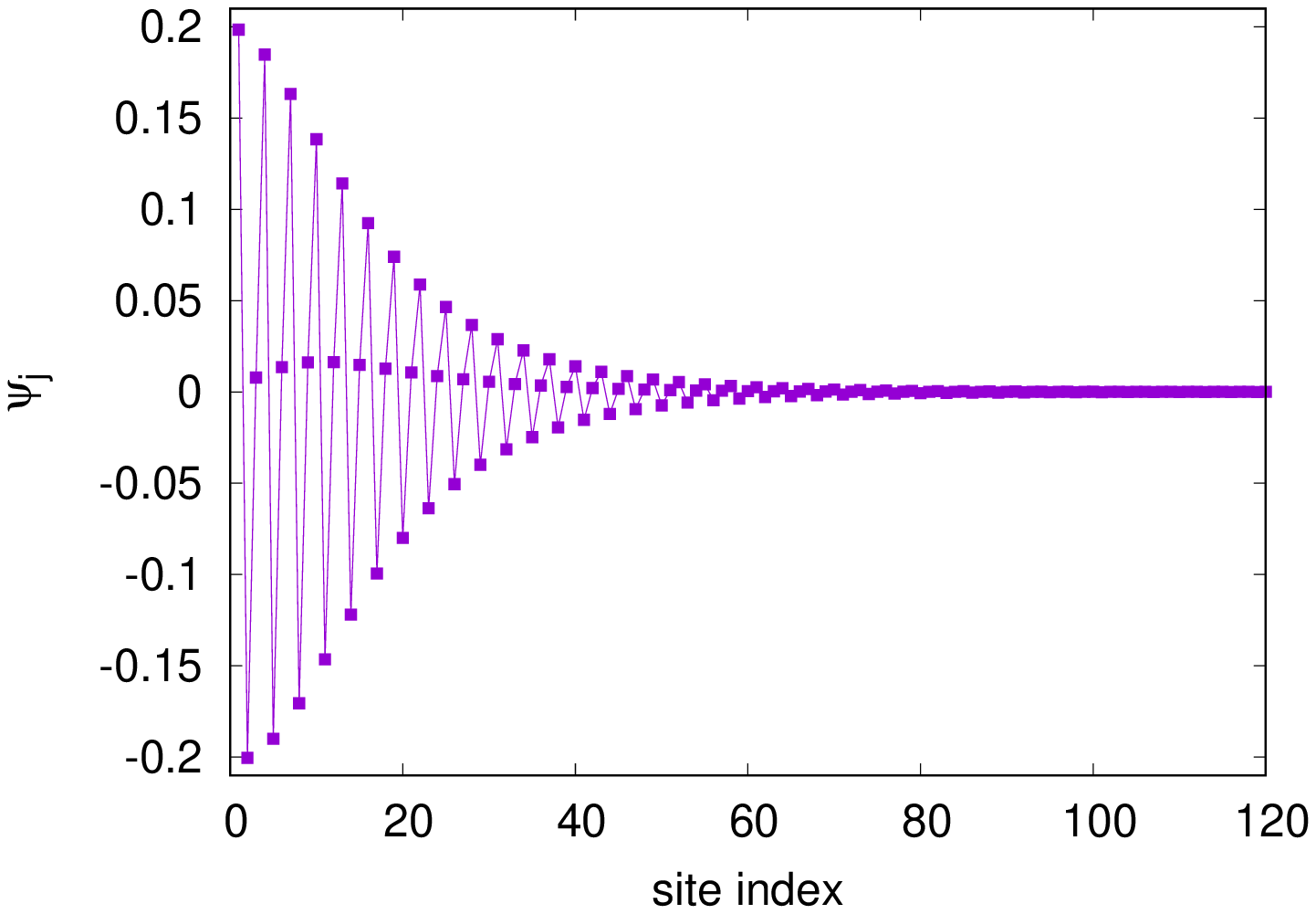}
 \caption{Upper: Norm versus frequency at fixed $w=2$, $v=1$.
   Lower: Mode amplitudes close to the band edge, $\lambda=-1.98$.
   \label{fig:Norm}} 
 \end{figure}

 Performing the continuation versus frequency at fixed $v$ (here
 $v/w=0.5$) it traces out the full continuous family of localized edge modes
 in the lower gap region, with the transition to a non-localized
 ``out-gap'' mode with  non-decaying tail at the upper gap edge, and the
 approach to a small-amplitude continuum-like mode at the lower gap edge.
 As illustrated in the upper Fig.\ \ref{fig:Norm}, the norm
 $P=\sum_j|\psi_j|^2$ describes 
 a monotonously increasing function $P(\lambda)$, approaching zero at the lower
 gap edge ($\lambda=-2$) and diverging at the upper
 ($\lambda=-\sqrt{3}+1 \approx-0.732$.) Thus, the localized nonlinear edge mode
 appears as a bifurcation from the linear extended mode at the upper
 edge of the lower band, without excitation threshold.
 The shape of the mode close to the
 lower gap edge is illustrated in the lower Fig.\ \ref{fig:Norm}, and is
 clearly seen to have a continuous envelope modulation of the linear band
 edge mode,
 with the main-field amplitudes ($A$ and $C$ sites) approaching a maximum with
 zero derivative and the small-field amplitude ($B$ sites) approaching zero at
 the edge of the chain. Thus, the edge mode has the structure of a ``half''
 bulk gap
 soliton, cut in the middle by the edge boundary condition. This will be
 confirmed below by deriving appropriate modulation equations in a
 continuous approximation.

\subsection{Continuous gap soliton limit}
\label{sec:cont}
In a continuous approximation of (\ref{eq:dyn}) we replace the site index
$n$ with a continuous variable $x=n$, and look for stationary solutions of
the form $$(A(x),B(x),C(x))e^{i \lambda t},$$ where $A(x), B(x), C(x)$ are
slowly varying real-valued functions.
Then, to lowest order approximation we replace
$A_{n+1}\rightarrow A(x)+ dA(x)/dx$, $C_{n-1}\rightarrow C(x)- dC(x)/dx$,
turning Eqs.\ (\ref{eq:dyn}) into
\begin{eqnarray}
  -\lambda A+wC-w\frac{dC}{dx}+vB+A^3 =0 ,  \label{eq:cont1}\\
   -\lambda B+v(A+C)+B^3=0 ,  \label{eq:cont2}\\  
  -\lambda C+vB +wA+w\frac{dA}{dx}+C^3 =0 ,
  \label{eq:cont3}
\end{eqnarray}
with boundary condition $B(0)=0$.
Analogously to the three-atomic elastic chain analyzed in
Ref.\ \onlinecite{UKV95},
there are two first-order differential equations for the fields $A$ and $C$
combined with a local algebraic equation for the field $B$. (As noted
in Ref.\ \onlinecite{UKV95}, this structure also generalizes to $N$-mers
yielding
$N-2$ algebraic equations.) Here, as $B$ defines the small-amplitude field
we may to lowest order neglect the cubic term in (\ref{eq:cont2}) and obtain
$B\simeq \frac{v}{\lambda}(A+C)$.
Adding and subtracting Eqs.\ (\ref{eq:cont1}) and
(\ref{eq:cont3}) and defining new functions $f=(A-C)/2$, $\phi=(A+C)/2$ then
leads to the system

\begin{eqnarray}
  w \frac{d\phi}{dx}=f\left[(-w-\lambda)+f^2+3\phi^2\right] ,
  \label{eq:cont_gap1}\\
  w \frac{df}{dx}=\phi\left[(-w+\lambda-\frac{2v^2}{\lambda})
    -(\phi^2+3f^2)\right].  \label{eq:cont_gap2}
\end{eqnarray}
These equations are equivalent to the well known gap soliton equations
for a diatomic elastic chain
whose exact analytical solutions in terms of hyberbolic functions
were analyzed in detail by
Chubykalo, Kovalev and Usatenko\cite{CKU93}, and later extended
by Kovalev et al.\cite{KUG99} to
more general systems resulting in different coefficients in front of the
nonlinear terms as compared to (\ref{eq:cont_gap1})-(\ref{eq:cont_gap2}).
Here Eq.\ (\ref{eq:cont_gap1}) describes the spatial shape of the
small-amplitude field $\phi$ and (\ref{eq:cont_gap2}) that of the
large-amplitude field $f$. The boundary condition at the edge yields
$\phi(0)=0$, which through (\ref{eq:cont_gap2}) also
implies the boundary condition
$\frac{df}{dx}(0)=0$ for the main field $f$. Thus, the edge boundary
conditions here correspond exactly to the conditions at the center of a standard
bulk gap soliton, which explains the shape of the edge mode as a``half'' gap
soliton shown in Fig.\ \ref{fig:Norm}
(compare, e.g., with Fig.\ 2 in Ref.\ \onlinecite{CKU93}). The maximum
amplitude for the main field $f$ of the analytical gap soliton is
obtained\cite{CKU93} from the linear term in  (\ref{eq:cont_gap1}) as
$f_{max}=\sqrt{2(\lambda+w)}$, which for the parameter values
in Fig.\ \ref{fig:Norm} yields $f_{max}=0.2$, in very good agreement with the
numerical results. So from the solutions to
(\ref{eq:cont_gap1})-(\ref{eq:cont_gap2}) we obtain, assuming the
continuous approximation to be valid
close to the lower gap edge $\lambda \gtrsim -w$,
$A_n\approx \phi(x)+f(x)$, $B_n\approx-\frac{2v}{w}\phi(x)$,
$C_n\approx \phi(x)-f(x)$.

We should note that in deriving (\ref{eq:cont_gap1})-(\ref{eq:cont_gap2})
we made slightly different assumptions than in Refs.\
\onlinecite{CKU93,UKV95,KUG99}. (i) We did not assume the gap width (here
given by $(3w-\sqrt{w^2+8v^2})/2$) to be small. This limits the validity
of our continuous approximation to the regime close to the lower band
edge, $\lambda \gtrsim -w$, and in particular the transition to ``outgap''
edge soliton at the upper gap edge is generally not accurately described
by (\ref{eq:cont_gap1})-(\ref{eq:cont_gap2}). (ii) We neglected the cubic term
in the small-amplitude field $B$ in (\ref{eq:cont2}). A more
accurate approximation yields
$B=\frac{2v}{\lambda}\phi+\frac{8v^3}{\lambda^4}\phi^3 + {\mathcal O}(\phi^5)$. 
Including also the last term will shift the coefficient for the  $\phi^2$ term
in (\ref{eq:cont_gap2}) from unity to $1+8(\frac{v}{\lambda})^4$,
which does not change qualitatively the shape
of the solution since $\phi^2 << 3f^2$ when the soliton amplitude is
non-negligible (the effect of changing this coefficient
was discussed in Ref.\ \onlinecite{KUG99}). Quantitatively, we do however
observe some deviations between the predictions from the continuum model and
the numerically calculated exact edge modes, as concerns the amplitude and
location of the peak of the small-amplitude field $B$. This is likely due
to the neglect of higher derivatives in (\ref{eq:cont1})-(\ref{eq:cont3}).

\section{Linear stability}
\label{sec:stab}
Considering the stability of the family of nonlinear edge modes, we perform
a standard numerical
linear stability analysis
(cf.\ e.g.\ Refs.\ \onlinecite{GJ04},\onlinecite{VJ09}).
Expressing an exact stationary edge mode as
$\psi_j(t) = \psi_j^{(\lambda)} e^{i\lambda t}$, with $\psi_j^{(\lambda)}$
time-independent, the linear stability is determined 
by adding a small perturbation written in the form
$\psi_j(t) = [\psi_j^{(\lambda)}+\epsilon_j(t)] e^{i\lambda t}$, with
$\epsilon_j(t) =\frac{1}{2}(\xi_j+\eta_j) e^{-i\omega_l t}+
\frac{1}{2}(\xi_j^\ast-\eta_j^\ast) e^{i\omega_l^\ast t}$. Linearizing and
solving the resulting eigenvalue problem yields the small-amplitude oscillation
frequencies $\omega_l$, with linear stability requiring all $\omega_l$ to be
real.

 \begin{figure*}
   \includegraphics[height=0.32\textwidth,angle=270]{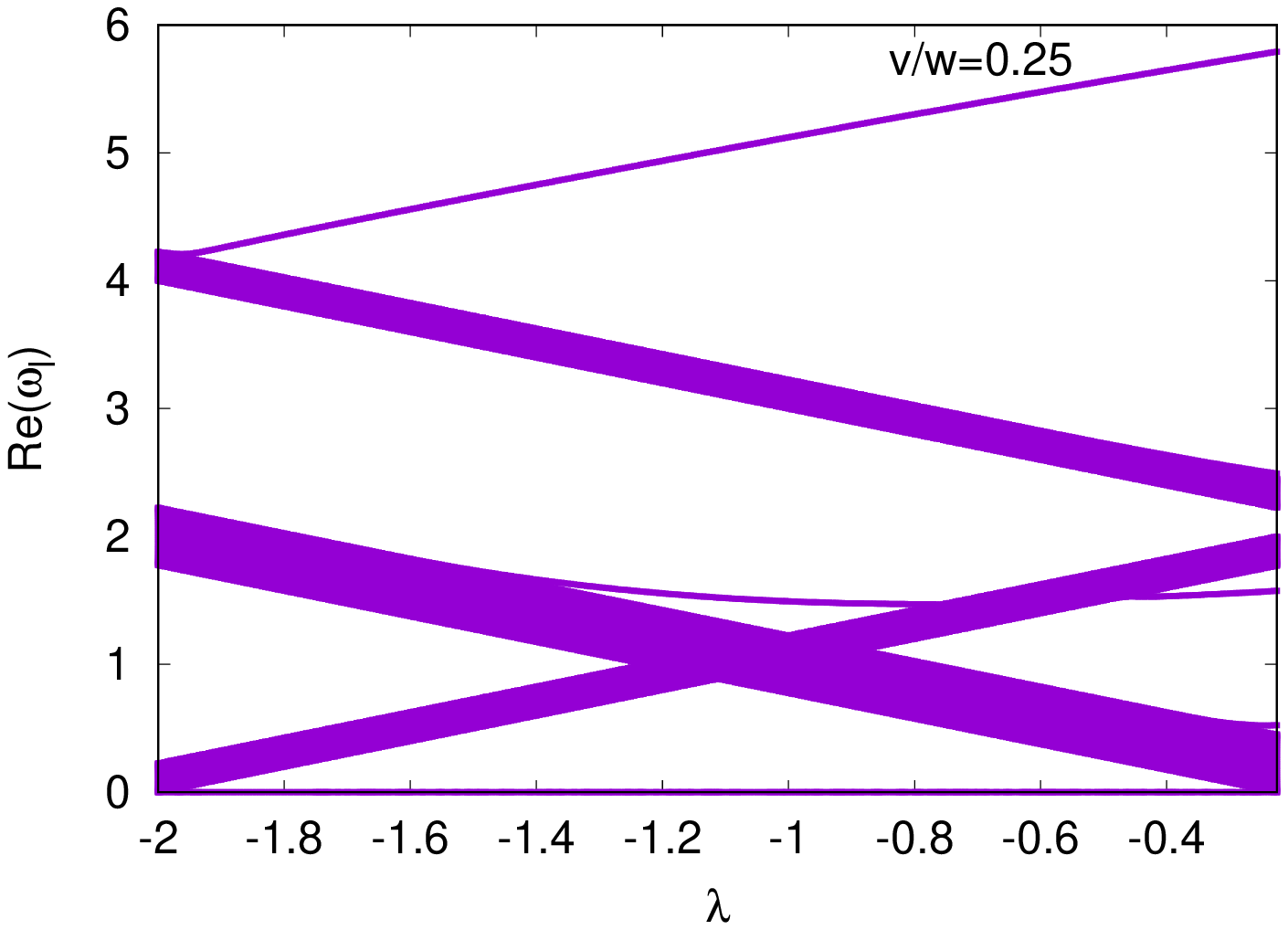}
   \includegraphics[height=0.32\textwidth,angle=270]{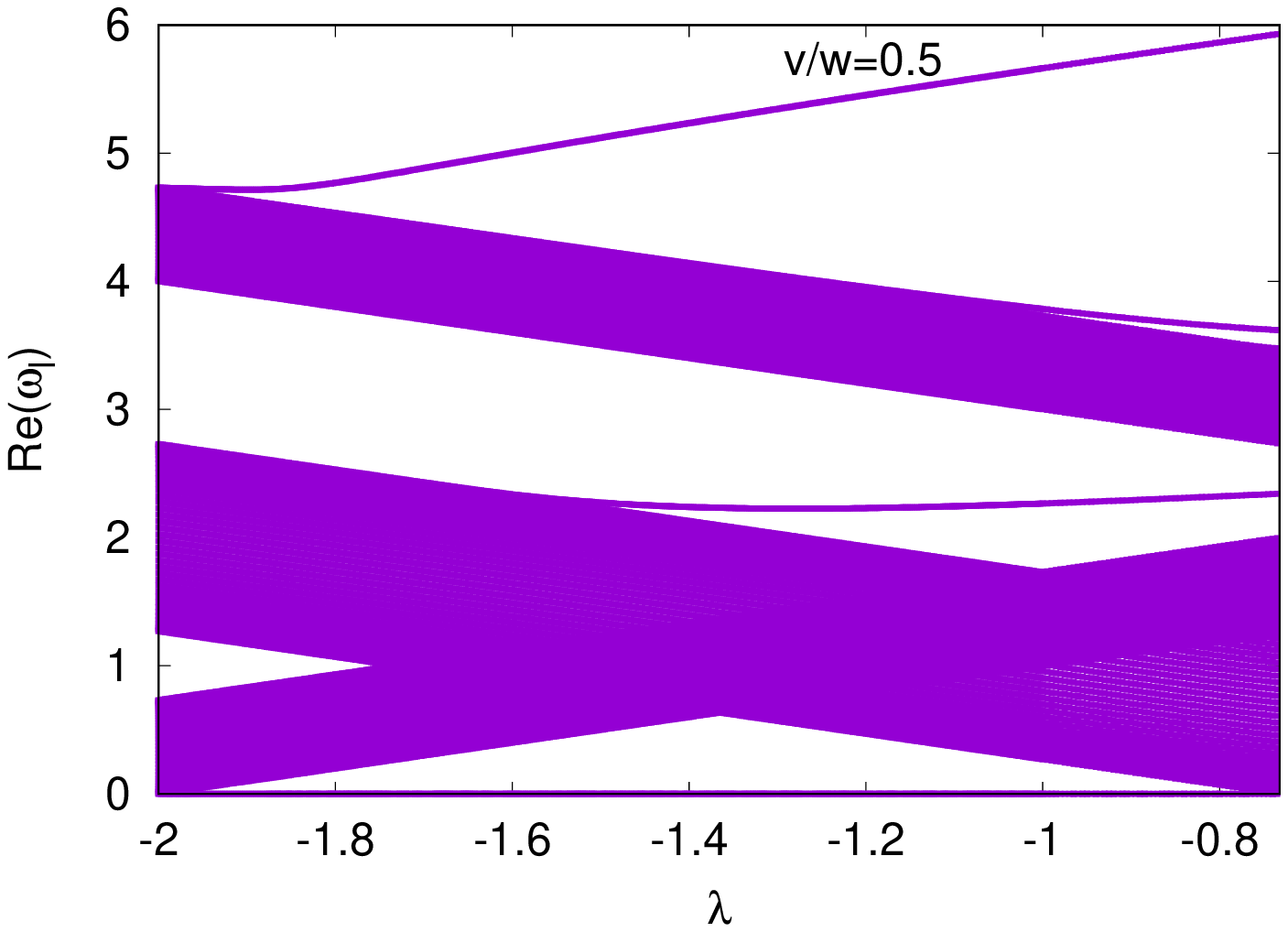}
   \includegraphics[height=0.32\textwidth,angle=270]{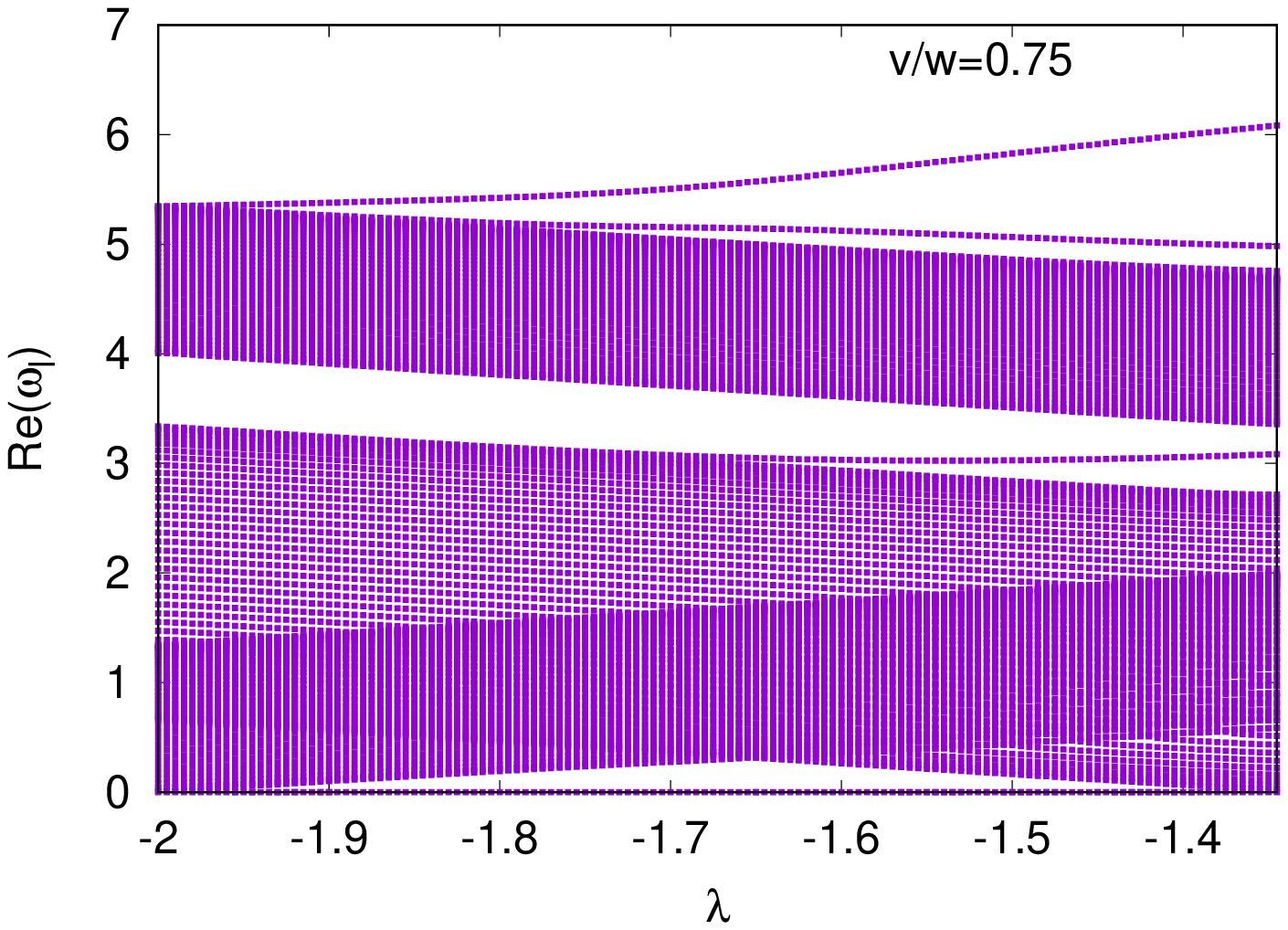}
   \includegraphics[height=0.32\textwidth,angle=270]{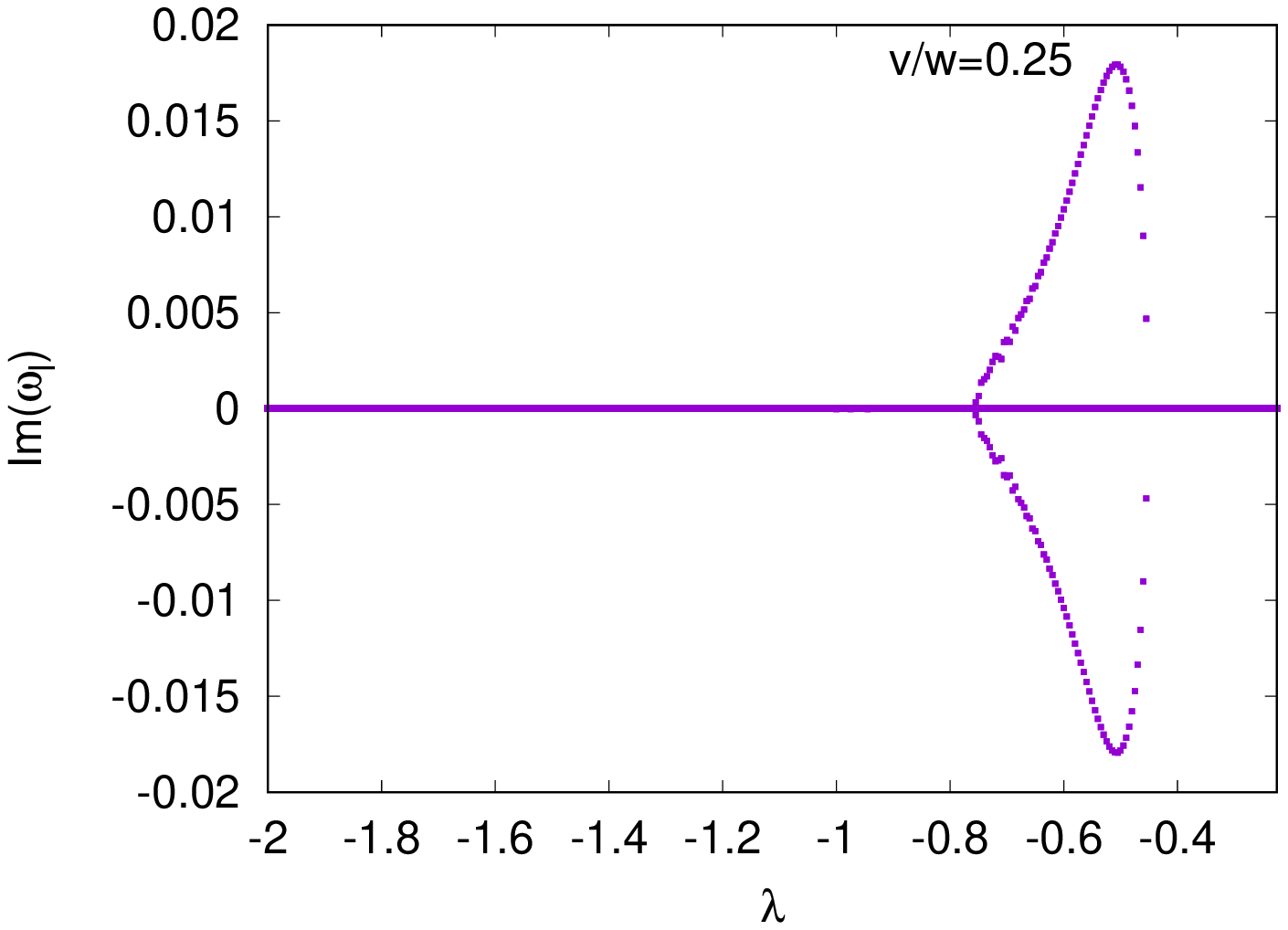}
   \includegraphics[height=0.32\textwidth,angle=270]{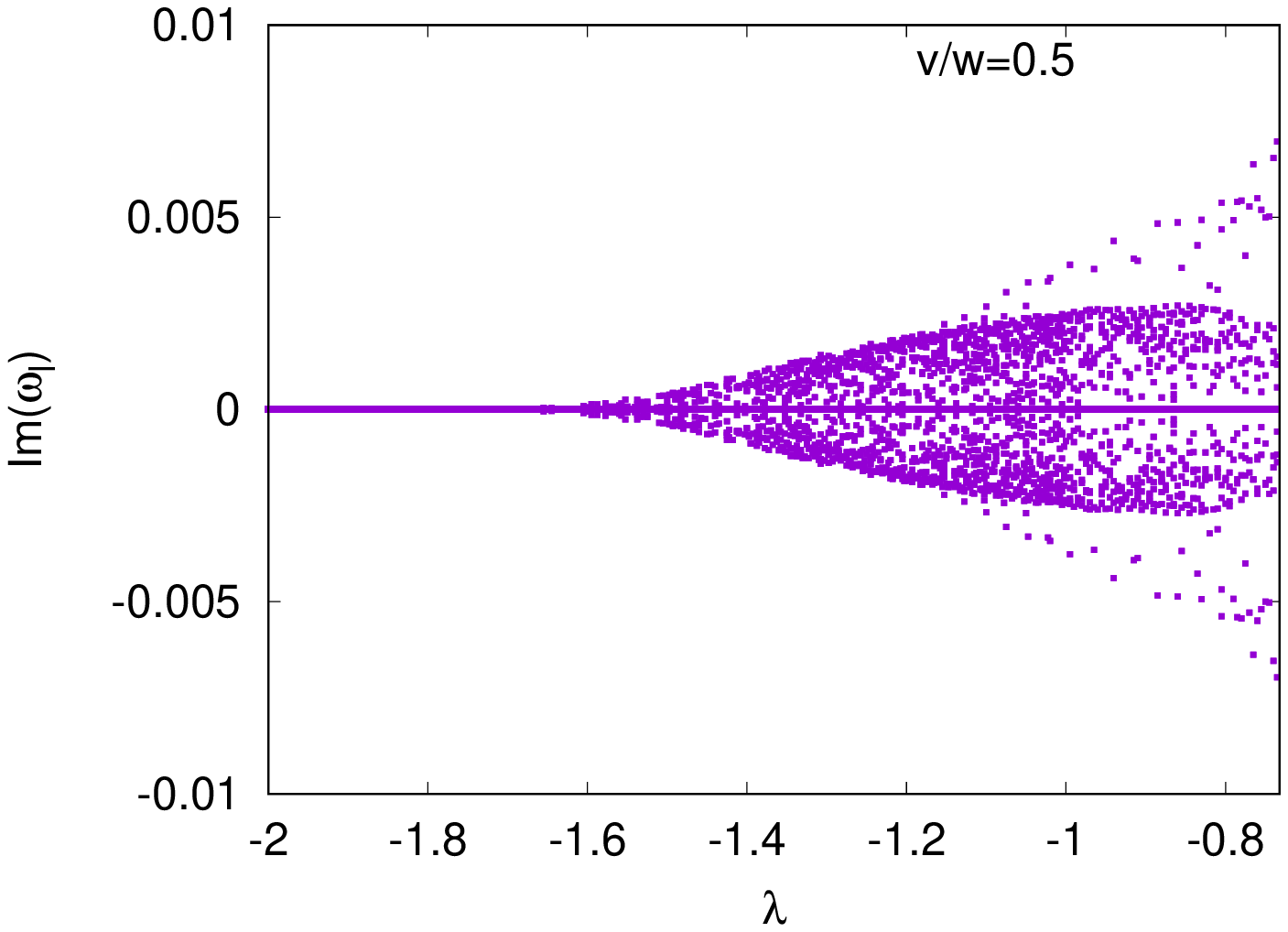}
   \includegraphics[height=0.32\textwidth,angle=270]{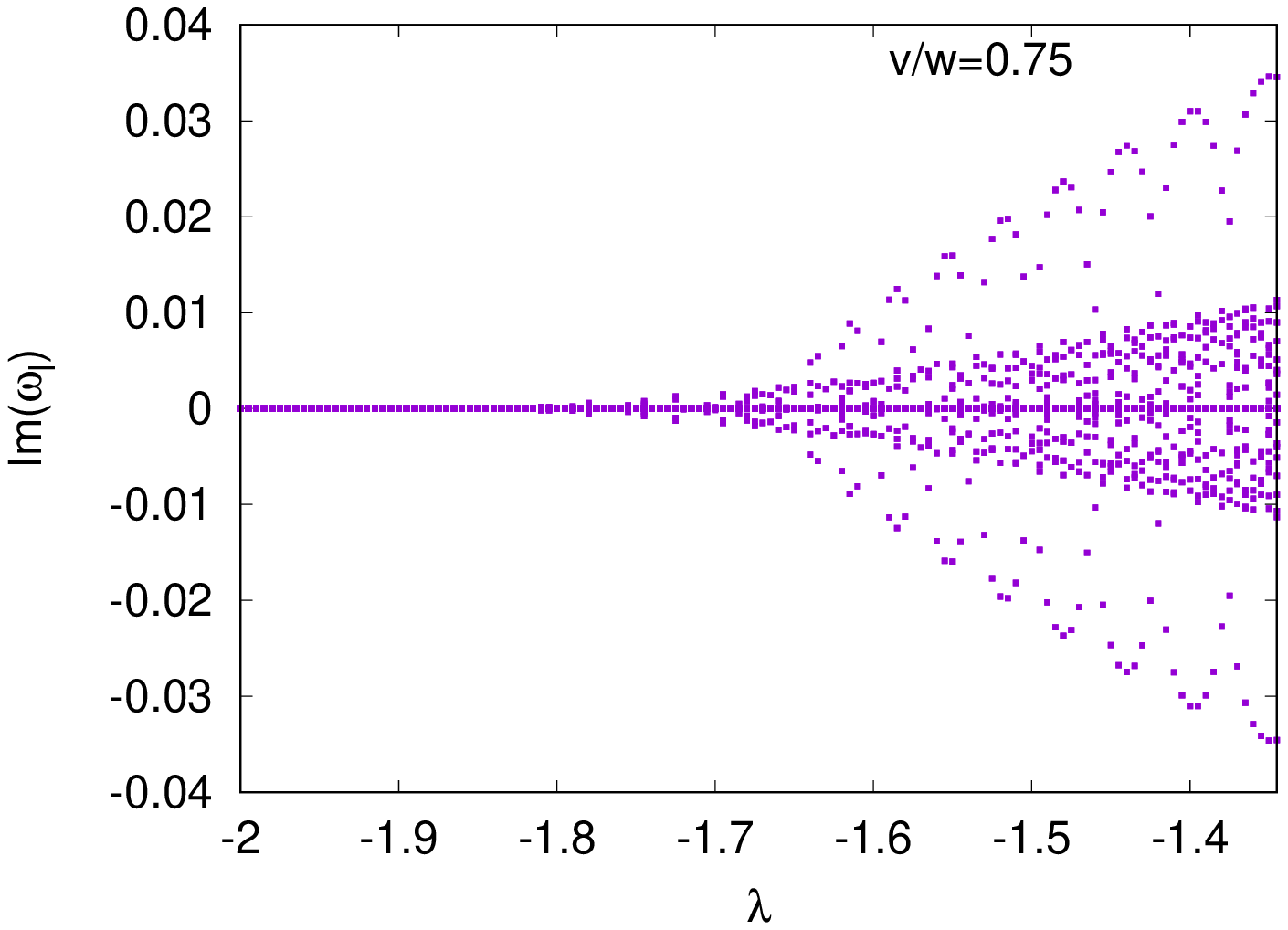}   
   \caption{Real and imaginary parts of the linear stability eigenfrequencies 
     for three values of $v/w$ as indicated in the figures, with $w=2$ and
     $N=40$ unit cells. Only the upper half (positive real part) of the
     spectrum is shown in the top plots. 
   \label{fig:stab}} 
 \end{figure*}
The linear stability properties for three different values of $v/w$ are
illustrated in Fig.\ \ref{fig:stab}. Qualitatively, the pictures show strong
resemblance with the corresponding stability properties for bulk gap solitons
in the standard (dimer) SSH model \cite{VJ09}.
As seen from the bottom plots in Fig.\ \ref{fig:stab}, the edge modes are
always linearly stable in the lower part of the gap (small-amplitude regime),
while oscillatory instabilities (complex eigenvalues) generally develop in
the upper part. By comparing with the top plots for the real parts of
oscillation frequencies, we see that the stronger instabilities appear
when isolated eigenvalues, bifurcating from some subband and
corresponding to  localized internal mode oscillations, resonate with
band eigenvalues (extended eigenmodes corresponding to a continuous
spectrum in the limit of infinite system size) from some other subband. The
broad spectrum of very weak instabilities is a standard signature of
overlapping continuous spectra, yielding finite-size instabilities
with strength decaying
towards zero as system size increases. It is also interesting to note that for
smaller values of $v/w$ the edge mode regains stability close to the upper
gap edge, while the instability  persists for larger $v/w$. Similar features
are seen both for bulk gap \cite{VJ09} and edge \cite{MS21,Bugarski} solitons
in the dimer SSH model.

Another interesting feature seen in Fig.\ \ref{fig:stab},
analogous to bulk gap solitons in dimer SSH,
is the presence of a localized mode with purely real eigenfrequency
above all continuous bands. In fact, in the limit $v\rightarrow 0$ this mode
is the same symmetric eigenmode ($\epsilon_1=\epsilon_2$) found in
Ref.\ \onlinecite{VJ09},
now localized at the edge sites and having eigenfrequency
$\omega_l=2w\sqrt{\frac{\lambda}{w}+2} + {\mathcal O}(v^2)$. 
As the upper edge of the spectrum of extended eigenmodes for small $v$ is
located at $\omega_l=w-\lambda + {\mathcal O}(v^2)$, it follows directly
that this
eigenmode is located above all continuous bands for all $\lambda$ in the gap
when $v$ is small. As the numerical results in Fig.\ \ref{fig:stab}
indicate, the mode indeed stays above the bands also for increasing $v$. Thus,
as also all higher harmonics of this eigenmode will reside above the
continuous spectrum, we expect that also ``edge breathers'' with
time-periodically oscillating intensity will exist as exact, non-radiating
solutions. Such edge breathers were explicitly found for the dimer SSH model
in Ref.\ \onlinecite{MJ23}.
An essential difference is that edge breathers in the dimer SSH result from
eigenmodes in the gap, which make them more prone to instabilities and
resonances. The edge breathers in the trimer SSH model, resulting from an
eigenmode above the continuous spectrum, are expected to exist stably in
a larger domain, more similar to the ``pulsons'' resulting from a similar
eigenmode for bulk gap solitons in the model with binary modulated on-site
potential \cite{JG04}.



%

\section{Conclusions}
In conclusion we presented a model system, a trimer array (SSH3) starting
with a single
strong bond followed by two weak bonds, which does not support localized
edge modes in the linear limit but where an arbitrarily weak on-site
nonlinearity is sufficient to induce a thresholdless localized
nonlinear edge mode,
without explicitly breaking the symmetry of the chain. We showed that,
close to the gap edge from which it bifurcates,  the mode
envelope could be approximated by exactly one half of a continuous bulk gap
soliton. The linear stability of the edge mode also indicated a similar
scenario as for bulk gap solitons in the standard SSH model, with stability for
smaller amplitude  and regimes of oscillatory instabilities appearing with
increasing amplitude. We also pointed out that conditions for existence of
time-periodically oscillating, localized edge breathers are generically
fulfilled for the full range of parameter values. These phenomena may be
experimentally observed e.g.\ with setups similar to those used in
Ref.\ \onlinecite{Kartashov22}.

%

%

 \begin{acknowledgments}
   It is a pleasure to contribute this work to the celebration of
   the 80th birthday of
   Alexander Kovalev, among many other things pioneer in the study of
   gap solitons in nonlinear lattices. Our joint collaboration projects had a
   strong impact on the
   activities of our group, and I have highly appreciated his deep knowledge
   in nonlinear physics, as well as his personal kindness and generosity.
 \end{acknowledgments}




\end{document}